\documentclass[aps,pra,longbibliography,twocolumn,superscriptaddress]{revtex4-1}

\usepackage{graphicx}
\usepackage{amsmath,amssymb}
\usepackage{txfonts}
\usepackage{bm}
\usepackage{color}
\usepackage{hyperref}
\usepackage{booktabs}
\usepackage{here}
\begin{document}
\title{Electrically Benign Defect Behavior in Zinc Tin Nitride Revealed from First Principles}
 
\author{Naoki Tsunoda}
\affiliation{Laboratory for Materials and Structures,  Institute of Innovative Research, Tokyo Institute of Technology, Yokohama 226-8503, Japan}

\author{Yu Kumagai}
\email[]{yuuukuma@gmail.com}
\affiliation{Materials Research Center for Element Strategy, Tokyo Institute of Technology, Yokohama 226-8503, Japan}
\affiliation{PRESTO, Japan Science and Technology Agency, Tokyo 113-8656, Japan}

\author{Akira Takahashi}
\affiliation{Laboratory for Materials and Structures,  Institute of Innovative Research, Tokyo Institute of Technology, Yokohama 226-8503, Japan}

\author{Fumiyasu Oba}
\affiliation{Laboratory for Materials and Structures,  Institute of Innovative Research, Tokyo Institute of Technology, Yokohama 226-8503, Japan}
\affiliation{Materials Research Center for Element Strategy, Tokyo Institute of Technology, Yokohama 226-8503, Japan}
\affiliation{Center for Materials Research by Information Integration, Research and Services Division of Materials Data and Integrated System, National Institute for Materials Science, Tsukuba 305-0047, Japan.}

\date{\today}

\begin{abstract}
Zinc tin nitride (ZnSnN$_2$) is attracting growing interest as a non-toxic and earth-abundant photoabsorber for thin-film photovoltaics.
Carrier transport in ZnSnN$_2$ and consequently cell performance are strongly affected by point defects with deep levels acting as carrier recombination centers.
In this study, the point defects in ZnSnN$_2$ are revisited by careful first-principles modeling based on recent experimental and theoretical findings. 
It is shown that ZnSnN$_2$ does not have low-energy defects with deep levels, in contrast to previously reported results. Therefore, ZnSnN$_2$ is more promising as a photoabsorber material than formerly considered.
\end{abstract}

\maketitle

ZnSnN$_2$ has a wurtzite-derived structure with a minimum gap of 1.4 eV in its direct-type band structure~\cite{bm14_2013,lany_2017}, which is very close to the best place of the Shockley--Queisser limit of 1.34 eV~\cite{shockley}. 
In addition, the electron effective mass of 0.17{\it m$_0$} and its heavy-hole mass of 2.00{\it m$_0$} are comparable or even superior to those of GaN (Fig. S2, Supplemental Material~\footnote[1]{See Supplemental Material at [URL will be inserted by publisher] for details of bulk and defect properties in ZnSnN$_2$.})~\cite{Hinuma:2016ie}. 
However, there are several fundamentally and technologically important issues regarding the use of ZnSnN$_2$ as a light absorption layer that need to be assessed: (i) {\it Point defect properties.} 
It is necessary to identify any point defects with deep levels, because they trap electrons and/or holes and cause carrier recombination, leading to loss of cell efficiency. (ii) {\it Unintentional n-type doping.} 
The carrier-electron concentration generally unintentionally increases up to 10$^{21}$ cm$^{-3}$ in ZnSnN$_2$ as well as other narrow gap nitrides like InN, Zn$_3$N$_2$, and ScN~\cite{doi:10.1063/1.1704853,PhysRevApplied.8.014015,PhysRevApplied.9.034019}. 
For solar cell applications, the carrier-electron concentration needs to be lowered to around 10$^{16}$-10$^{18}$ cm$^{-3}$. 
Besides, the photoabsorption onset is increased up to 2.4 eV~\cite{bm14_2013,feldberg_2013,veal_2015} because of the Burstein--Moss (BM) shift~\cite{PhysRev.93.632,0370-1301-67-10-306}, which has raised considerable controversy regarding the fundamental gap of ZnSnN$_2$. 
The BM shift is also directly related to the formation of donor-type point defects.
Understanding the point-defect properties of ZnSnN$_2$ is thus imperative to optimize its performance in photovoltaics. 

Chen {\it et al.}~\cite{chen_2014} investigated native point defects and oxygen impurities in ZnSnN$_2$ with an ordered orthorhombic structure.
They reported that Sn-on-Zn antisite (Sn$_{\rm Zn}$) and O-on-N (O$_{\rm N}$) substitutional defects have low formation energies with deep donor levels, and thus do not cause the BM shift in ZnSnN$_2$.
However, given recent experimental and theoretical findings~\cite{bm14_2013,lany_2017,feldberg_2013,veal_2015,fioretti_2015,opd_2015,hydrogen_2017}, this conclusion needs to be reviewed. This motivated us to reevaluate point defects in ZnSnN$_2$ including as-yet-uninvestigated hydrogen impurities.
Our results show that all the defects with deep levels are very high in energy and the abundant defects act as shallow donors.
Considering these favorable defect properties together with its appropriate band gap and high absorption coefficient, ZnSnN$_2$ is a promising photoabsorber for thin-film photovoltaics.

Our theoretical investigation of the point defects in ZnSnN$_2$ was performed for the {\it Pna2$_1$} orthorhombic structure with 16 atoms in the unit cell, referred to here as the ordered model.
However, a certain level of disorder in the cation sublattice appears to be unavoidable because of the very low cation order--disorder transition temperature~\cite{lany_2017}.
At low temperature, the disordered phase strictly retains the local charge neutrality, in which each N atom is necessarily coordinated by two Sn and two Zn atoms.
We refer to such a structure as the disordered structure with local charge neutrality (DLCN).
It has been reported that the electronic structure of the DLCN is almost identical to that of the ordered structure~\cite{lany_2017,opd_2015}.
Thus, the DLCN should be ideal as a photoabsorber, unless its defects are detrimental to its efficiency.
We thus considered both ordered and DLCN models in our defect calculations.
Note that cation disorder breaking the local charge neutrality occurs at very high temperature ($>$ 1750~K~\cite{lany_2017}).
Some authors have used special quasirandom structures (SQS) to examine the behavior of the cation disordered phase~\cite{feldberg_2013,veal_2015}. Such a fully random cation disordered model is not appropriate for ZnSnN$_2$ because its typical growth temperature is much lower than 1750~K~\cite{bm14_2013,feldberg_2013,veal_2015,fioretti_2015,hydrogen_2017}.
Besides, when the charge neutrality is broken, a high concentration of Zn-on-Sn (Zn$_{\rm Sn}$) defects is introduced.
This situation should be avoided for photovoltaic applications because Zn$_{\rm Sn}$ defects give rise to deep levels, as discussed later.

Figure 1 shows the two models used to evaluate defect formation energies.
The DLCN model was generated by Monte Carlo simulated annealing of a 128-atom orthorhombic supercell.
The DLCN model shown in Fig. 1 belongs to space group {\it Pna2$_1$} and consists of a 32-atom unit cell, consistent with that reported elsewhere~\cite{lany_2017}. 
Note that there are infinite possible configurations for the DLCN models when different cell sizes are considered; however, within the 128-atom supercell, we found only one configuration owing to the strong geometrical constraint of the local charge neutrality.

First, we discuss the band gap of ZnSnN$_2$ because it is a central controversy in the research of this material.
As the carrier-electron concentration increases, the photoabsorption onset also increases due to the BM shift. The blueshift estimated from the Heyd-Scuseria-Ernzerhof (HSE06) hybrid functional calculations is 0.2-2.3 eV when the carrier-electron concentration is 10$^{18}$-10$^{21}$ cm$^{-3}$ observed in experiments~\cite{bm14_2013,feldberg_2013,veal_2015} (Fig. S6, Supplemental Material~\footnotemark[1]).
Thus, larger experimental band gaps than 1.4 eV could be attributed to the BM shift, as has also been discussed by Lahourcade et al.~\cite{bm14_2013}
Second, the band gaps of fully cation disordered systems are ill-defined because the minimum gap decreases as the model size increases~\cite{sqs_seko}.
In the case of ZnSnN$_2$, the density of states tends to continuously develop within the band gap when N atoms are coordinated by three or four Sn or Zn atoms, as discussed later.
Note that this does not necessarily indicate metallic behavior because the defect states are discontinuous in real space.
Refer to the Supplemental Material for a more detailed discussion considering SQS (Fig. S8~\footnotemark[1]).

\begin{figure}
  \includegraphics[width=1\linewidth]{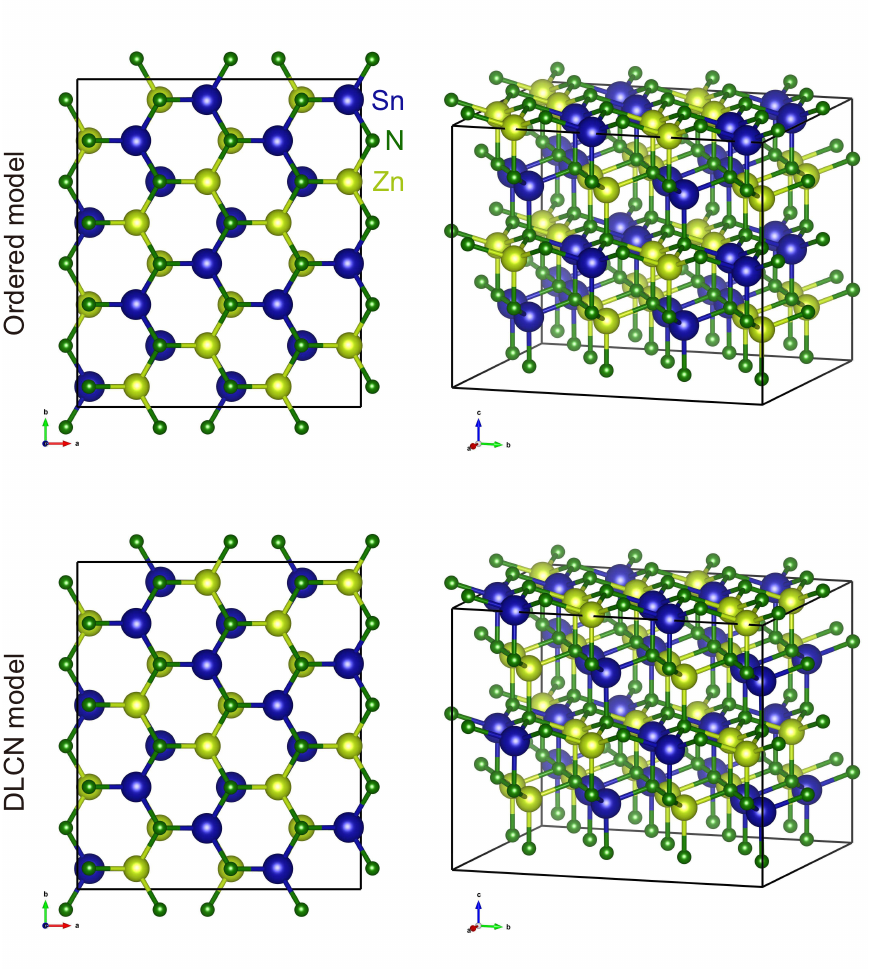}
  \caption{Atomic structures of the ordered and DLCN models of ZnSnN$_2$ used in this study. These models contain 128 atoms. Both structures belong to space group {\it Pna2$_1$}.}
  \label{structures_fig}
\end{figure}

\begin{figure*}[t]
 \begin{center}
  \includegraphics[width=1\linewidth]{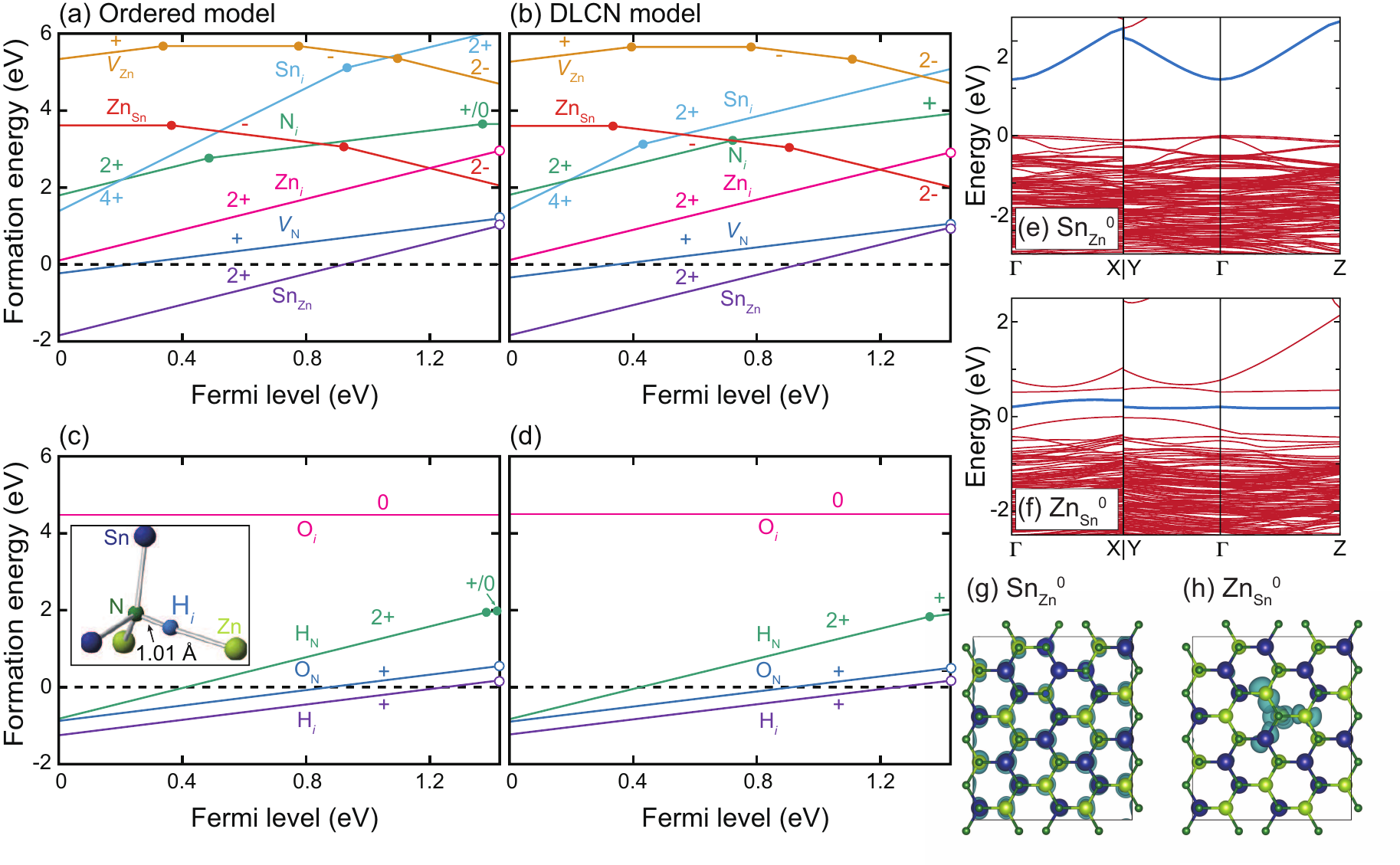}
   \caption{Formation energies of point defects in ZnSnN$_2$ as a function of the Fermi level. (a, b) Native defects and (c, d) impurities in the (a, c) ordered model and (b, d) DLCN model. The valence band maximum is set to the zero of the Fermi level and the upper limit corresponds to the conduction band minimum. The defect species and sites are indicated by {\it X$_Y$}, where $X$ means a vacancy ($V$) or an added element and $Y$ is an interstitial site ($i$) or substitutional site. The chemical potentials are set at a point where the Zn and Sn metals equilibrate with ZnSnN$_2$. Note that shallow donor levels in the vicinity of the conduction-band minimum are designated by open circles (see text in detail). Note also that the Sn vacancy does not appear in Figures (a) and (b) owing to its very high formation energy (see Fig. S12, Supplemental Material). The inset of (c) shows the atomic structure for H$_i^{+}$. (e, f) Band structures for the Sn$_{\rm Zn}^{0}$ and Zn$_{\rm Sn}^{0}$ in the 128-atom supercell of the ordered model. Note that Zn$_{\rm Sn}^{0}$ shows spin polarization, and only the down-spin channel, which indicates two unoccupied deep localized states, is shown. (see Fig. S14, Supplemental Material, for the up-spin channel) (g, h) Squared wave functions of the highest occupied and the lowest unoccupied bands for Sn$_{\rm Zn}^{0}$ and Zn$_{\rm Sn}^{0}$, respectively, highlighted by blue color in Figures (e) and (f).
}
  \label{energies_fig}
 \end{center}
\end{figure*}

We now discuss the energetics of point defects in ZnSnN$_2$.
To investigate the site dependency in the DLCN model, we conducted over 300 point-defect calculations using the modified Perdew--Burke--Ernzerhof generalized gradient approximation tuned for solids (PBEsol-GGA)~\cite{PhysRevLett.100.136406}.
We confirmed that most of the defect species show only negligible site dependencies in energy, while the interstitial defects do relatively large ones probably due to the difference in neighboring cation configurations at the octahedral sites (Fig. S9, Supplemental Material~\footnotemark[1]).
Thus, we performed HSE06 calculations for a few configurations with lower energies in PBEsol calculations for each defect and show only the results of the lowest energy configurations hereafter.
Figure 2 shows the resultant formation energies of the native defects as well as hydrogen and oxygen impurities, which are ubiquitous in nitrides.
Since ZnSnN$_2$ is only slightly stable ($\Delta$$H$$_f$ = -0.015 eV/atom) compared to Zn metal, Sn metal, and N$_2$ molecules using HSE06, the growth condition does not largely affect the formation energies under the equilibrium condition.
Therefore, here, the chemical potentials of constituent elements are set at the point where Zn and Sn metals equilibrate with ZnSnN$_2$.
As mentioned above, the standard HSE06 functional used in this study yielded a direct gap of 1.4 eV.
In the study by Chen {\it et al.}, the Fock exchange ratio was increased to reproduce a reported experimental gap of 2.0 eV~\cite{bm14_2013} although one of the authors attributed this larger gap to the BM shift and stated that the fundamental gap is about 1.4 eV~\cite{bm14_2013}.

Before discussing the results, let us consider the technical details that can alter conclusions even qualitatively.
Chen {\it et al.}~\cite{chen_2014} did not consider the finite cell-size effects when calculating defect formation energies and the reciprocal space sampling was performed using only the $\Gamma$ point.
According to our calculated cell-size dependencies, a maximum error of 0.72 eV arises when the 128-atom supercell is used with these settings.
In this study, we decreased these errors using the extended Freysoldt--Neugebauer--Van de Walle (FNV) corrections~\cite{PhysRevB.89.195205,Freysoldt:2009ih} and 2$\times$2$\times$2 Monkhorst--Pack $k$-point sampling for the 128-atom supercell.
Consequently, the errors based on our test calculations are less than 0.06 eV.
In the case of defects exhibiting hydrogenic states, a huge supercell including tens of thousands of atoms is usually required to avoid overlap between widespread defect orbitals~\cite{PhysRevLett.110.166404}.
When using the 128-atom supercell in ZnSnN$_2$, the thermodynamic transition levels of such defects are overestimated by a few tenths of eV because of the defect--defect interaction (see Fig. S10, Supplemental Material~\footnotemark[1]).
Therefore, we discuss the transition levels associated with hydrogenic states only qualitatively.

Figures 2(a--d) reveals that there is little difference between the defect formation energies of the ordered and DLCN models, which means that not only the bulk properties but also the point-defect properties are strongly correlated with its immediate coordination environment.
This conclusion would also hold for similar cation disordered systems such as ZnSnP$_2$~\cite{PhysRevB.90.125202,zsp_scanlon} and ZnGeN$_2$~\cite{PhysRevB.93.155202} if the local charge neutrality is preserved.
It is also notable that defects and impurities that show low formation energies simultaneously with deep levels do not exist when the Fermi level is located within the band gap. This is in stark contrast to the conclusion of a previous study~\cite{chen_2014} that the Zn interstitial (Zn$_i$), Sn$_{\rm Zn}$, nitrogen vacancy ($V_{\rm N}$), and O$_{\rm N}$ have deep defect levels and act as carrier recombination centers.
This discrepancy is mainly caused by the aforementioned difference in the treatment of the cell-size corrections and $k$-point sampling. 

As shown in Figures 2(e) and (g), Sn$_{\rm Zn}^{0}$ exhibits an occupied hydrogenic state, namely perturbed host state, while the localized defect state is not confirmed near the conduction-band minimum (CBM), which is different from the results by Chen {\it et al.}~\cite{chen_2014}
Therefore, Sn$_{\rm Zn}$ is a dominant shallow donor among the native defects.
Its formation energy becomes zero at the Fermi level being 0.9 eV above the valence-band maximum (VBM), which hinders $p$-type conversion even with acceptor doping.
$V_{\rm N}$ is a single shallow donor in the entire Fermi level range, but has a higher formation energy than that of Sn$_{\rm Zn}$.
In contrast, acceptor-type defect Zn$_{\rm Sn}$ forms two deep transition levels between the 0, -1, and -2 charge states.
Indeed, as seen in Figures 2(f) and (h), Zn$_{\rm Sn}^{0}$ shows two deep localized defect states within the band gap.
However, the formation energy of Zn$_{\rm Sn}$ is rather high when the Fermi level is within the band gap.
The other defects, namely, Zn vacancy ($V_{\rm Zn}$), Sn vacancy ($V_{\rm Sn}$), Sn interstitial (Sn$_i$), and N interstitial (N$_i$), create deep levels but have very high formation energies.
It is also noteworthy that Sn$_{\rm Zn}^{0}$ does not become a DX center ($V_{\rm Zn}$ + Sn$_i$) in our calculations unlike ZnSnP$_2$~\cite{PhysRevB.90.125202} probably because of the high formation energies of $V_{\rm Zn}$ and Sn$_i$ and/or the difference in crystal structures.

Oxygen and hydrogen impurities are energetically favorable at the N sites (O$_{\rm N}$) and interstitial sites (H$_i$), respectively.
These impurities also act as single shallow donors and exhibit low formation energies even when the Fermi level is at the CBM.
Orbital analyses indicated that their donor electrons are located at the perturbed conduction bands.
Therefore, these impurities should primarily cause the BM shift.
However, in the previous study~\cite{chen_2014}, the origin of the high carrier-electron concentration was attributed to the defect band originating from Sn$_{\rm Zn}$ and O$_{\rm N}$, which is different from our results.
H$_i$ forms an N-H bond (1.01 \AA) between N and Zn atom, as commonly observed for nitrides. 
We also found that anionic hydrogen H$_i^{-1}$ is not stabilized even when the Fermi level is located at 1 eV above the CBM in ZnSnN$_2$ (see Fig. S12, Supplemental Material~\footnotemark[1]), unlike in GaN~\cite{PhysRevLett.75.4452}.

Recently, Fioretti {\it et al.}~\cite{hydrogen_2017} showed that annealing Zn-rich Zn$_{\rm 1+x}$Sn$_{\rm 1-x}$N$_2$ grown in a hydrogen atmosphere decreased its carrier-electron concentration to 4 $\times$ 10$^{16}$ cm$^{-3}$.
They explained this observation from the viewpoint of hydrogen passivation of acceptors, i.e., Zn$_{\rm Sn}$ + H$_i$ in Zn$_{\rm 1+x}$Sn$_{\rm 1-x}$N$_2$ during growth, which lowers the driving force for the formation of other unintentional donors.
Indeed, our results indicate that complexing with hydrogen is exothermic and drastically decreases the formation energy of acceptor Zn$_{\rm Sn}$ (Fig. S13~\footnotemark[1]); the binding energy, i.e., the energy change from isolated Zn$_{\rm Sn}$$^{-2}$ and H$_i^{+}$ to (Zn$_{\rm Sn}$ + H$_i$)$^{-}$ is -1.43 eV and that from isolated H$_i^{+}$ and (Zn$_{\rm Sn}$ + H$_i$)$^{-}$ to (Zn$_{\rm Sn}$ + 2H$_i$)$^{0}$ is -0.73 eV.
Therefore, abundant Zn$_{\rm Sn}$ antisites are easily introduced by hydrogen passivation as discussed by Fioretti {\it et al.}
We, however, emphasize that Zn$_{\rm Sn}$ antisites, which generate deep transition levels and trap minority carrier holes, persist even after removing the passivating hydrogen by post-deposition annealing.
Moreover, the determination of the intrinsic band gap would be inhibited by the optical absorption related to the defect band.

\begin{figure}
 \begin{center}
  \includegraphics[width=1\linewidth]{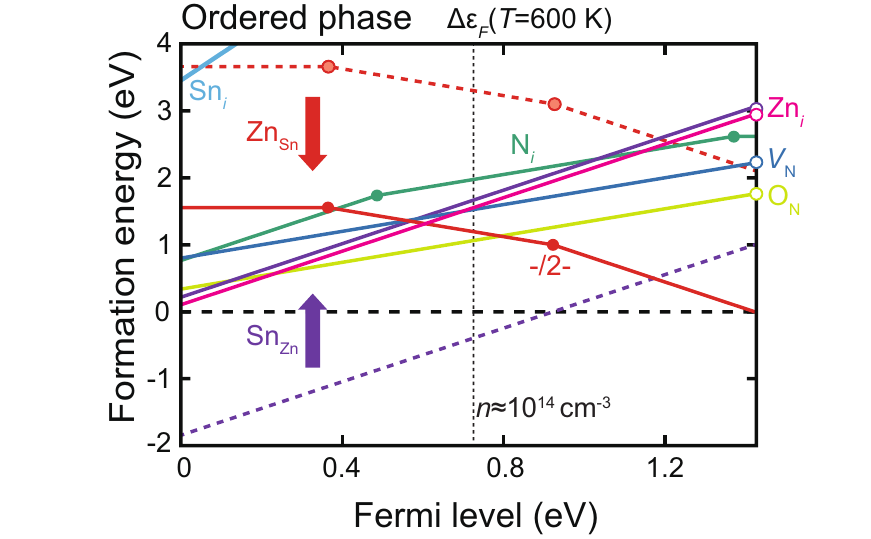}
  \caption{Same as Fig. 2, but the nitrogen chemical potential is increased by 1~eV under the Zn-rich condition (see text for details). Since hydrogen impurity could be reduced by post-growth annealing, we consider only oxygen as an impurity, here. For comparison, the dashed line shows the Sn$_{\rm Zn}$ and Zn$_{\rm Sn}$ formation energies under the Zn-rich and Sn-rich condition considered in Figure~\ref{energies_fig}.}
  \label{carrier_fig}
 \end{center}
\end{figure}

Here, we propose an alternative route to achieving ZnSnN$_2$ with lower carrier-electron concentration.
In the above discussion, the chemical potentials were set at a point where the Zn and Sn metals equilibrate with ZnSnN$_2$.
However, recent growth techniques such as reactive sputtering can be used to raise the chemical potential of N ($\mu_{\rm N}$) by up to +1 eV/N from that of an inactive N$_2$ molecule~\cite{metastable_nitride}. 
Consequently, metastable nitrides can be synthesized.
A notable example is Cu$_3$N, in which $\mu_{\rm N}$ needs to be at least +1.04 eV higher than that of an N$_2$ molecule~\cite{metastable_nitride}.
When $\mu_{\rm N}$ was increased to this value, we can use more advantageous condition for lowring the carrier-electron concentration.
Figure 3 shows the formation energies of native defects and oxygen impurity under $\Delta\mu_{\rm N}$ = +1 eV/N and Zn-rich (Sn-poor) condition (see Fig. S16 for details~\footnotemark[1]). 
The equilibrium Fermi level is located at the 0.70 eV from the CBM at 600 K which assumes synthesis temperature of ZnSnN$_2$~\cite{feldberg_2013,fioretti_2015}.
In this situation, all the defect concentrations are less than 10$^{14}$ cm$^{-3}$ and corresponding carrier concentration is $n$=1.9 $\times$ 10$^{14}$ cm$^{-3}$.
One might expect that $p$-type conversion by acceptor doping is plausible in this condition as all the donor-type defects compensating holes are high in energy.
Therefore, we performed the extensive calculations for impurities Li, Na, K, Cu, and C using HSE06. (see Fig. S17, Supplemental Material~\footnotemark[1])
As a result, however, we found that none of them act as effective acceptor dopants for ZnSnN$_2$ because of deep acceptor levels or incorporation into interstitial sites.
Thus, $p$-type conversion in ZnSnN$_2$ seems very difficult as previously considered even if compensating donor-type defects are sufficiently suppressed as in Fig. 3.~\footnote[2]{Note that Wang et al. previously investigated Li-, Na-, and K-on-Zn substitution as acceptors for ZnSnN$_2$, and concluded that Li-on-Zn substitution acts as a shallow acceptor~\cite{li_2017}. However, their study did not calculate the interstitial dopants which act as donors unlike our study.}

{\it Methods.} First-principles calculations were performed using the projector augmented wave (PAW) method~\cite{PhysRevB.50.17953,PhysRevB.59.1758} as implemented in VASP~\cite{PhysRevB.54.11169}. 
The PBEsol-GGA was used to calculate the total energies of various DLCN models and SQS with different sizes, and to check the site and cell-size dependencies of defect formation energies.
The HSE06 hybrid functional~\cite{JChemPhys.124.219906,JChemPhys.125.224106} with standard parameters was used for the calculations of the band structure, density of states, and defect formation energies. 
The plane-wave cutoff energy was set to 550 eV for the lattice relaxation of the ordered and DLCN models without defects and 400 eV for the other calculations with fixed lattice constants. 
The residual forces were converged to less than 0.01 and 0.04 eV/\AA\ for the calculations without and with defects, respectively.
Spin polarization was considered for all the defect calculations using HSE06.

The formation energy of a point defect was calculated as~\cite{PhysRevB.89.195205}
\begin{eqnarray}
E_f[D^q] &=& \Bigl\{E[D^q] + E_{\rm corr}[D^q]\Bigr\} - E_P +  \sum n_i\mu_i \nonumber \\
 &+& q (\epsilon_{\rm VBM}+ \Delta\epsilon_F),
 \label{defect_formation_energy}
\end{eqnarray}
where $E[D^q]$ and $E_P$ are the total energies of the supercell with defect $D$ in charge state $q$ and the supercell without a defect, respectively.
$n_i$ is the number of removed ($n_i$ $>$ 0) or added ($n_i$ $<$ 0) $i$-type atoms and $\mu_i$ is the chemical potential representing the growth conditions.
The referenced competing phases used were hexagonal Zn, cubic Sn, wurtzite ZnO, and N$_2$ and H$_2$ molecules.
$\epsilon_{\rm VBM}$ is the energy level of the VBM, and $\Delta\epsilon_F$ is the Fermi level ($\epsilon_F$) with respect to $\epsilon_{\rm VBM}$.
Therefore, $\epsilon_F = \epsilon_{\rm VBM} + \Delta\epsilon_F$. 
$E_{\rm corr}[D^q]$ corresponds to the correction energy for a finite supercell size error associated with spurious electrostatic interactions between charged defects.
We used our extended FNV correction scheme~\cite{PhysRevB.89.195205} in the calculations.
More computational details are described in the Supplemental Material.~\footnotemark[1]

Both DLCN and SQS models were generated by Monte Carlo simulated annealing using in-house and CLUPAN codes~\cite{PhysRevB.80.165122}, respectively (see Fig. S5, Supplemental Material, for more details~\footnotemark[1]).

{\it Conclusions.} We theoretically revisited the point defects in ZnSnN$_2$ by realistic modeling of its disordered phase with the local charge neutrality.
Our calculations revealed that the ordered and DLCN models exhibited nearly the same stability, volumes, electronic structures, and even point-defect properties, indicating these properties are determined mainly by the immediate coordination environment.
It was also found that low-energy defects with deep levels are absent in ZnSnN$_2$ and, therefore, there is less carrier recombination caused by point defects than thought previously.
Furthermore, a possible route to ZnSnN$_2$ with lower carrier-electron concentration using non-equilibrium growth techniques was proposed.
Our study has unveiled the further potential of ZnSnN$_2$ as a photoabsorber in thin-film photovoltaics.

\vspace{3mm}
{\it Acknowledgements.} This work was supported by the MEXT Elements Strategy Initiative to Form Core Research Center, Grants-in-Aid for Young Scientists A (Grant No. 15H05541) and 
Scientific Research A (Grant No. 17H01318) from JSPS, and PRESTO (JPMJPR16N4), and Support Program for Starting Up Innovation Hub MI$^2$I from JST, Japan.
The computing resources of ACCMS at Kyoto University were used for a part of this work.

%
\end{document}


\title{\Huge Supplemental Material}

\author[1]{Naoki Tsunoda}
\affil[1]{\small Laboratory for Materials and Structures,  Institute of Innovative Research, Tokyo Institute of Technology, Yokohama 226-8503, Japan}

\author[2,3]{Yu Kumagai}
\affil[2]{\small Materials Research Center for Element Strategy, Tokyo Institute of Technology, Yokohama 226-8503, Japan}
\affil[3]{\small PRESTO, Japan Science and Technology Agency, Tokyo 113-8656, Japan}

\author[1]{Akira Takahashi}

\author[1,2,4]{Fumiyasu Oba}
\affil[4]{\small Materials Research Center for Element Strategy, Tokyo Institute of Technology, Yokohama 226-8503, Japan}
\affil[5]{\small Center for Materials Research by Information Integration, Research and Services Division of Materials Data and Integrated System, National Institute for Materials Science, Tsukuba 305-0047, Japan.}

\date{\today}


\definecolor{blue}{rgb}{0.0,0.0,1}
\hypersetup{colorlinks,breaklinks,
            linkcolor=blue,urlcolor=blue,
            anchorcolor=blue,citecolor=blue}

\maketitle
\clearpage

\begin{figure}
  \centering
  \includegraphics[width=0.85\linewidth]{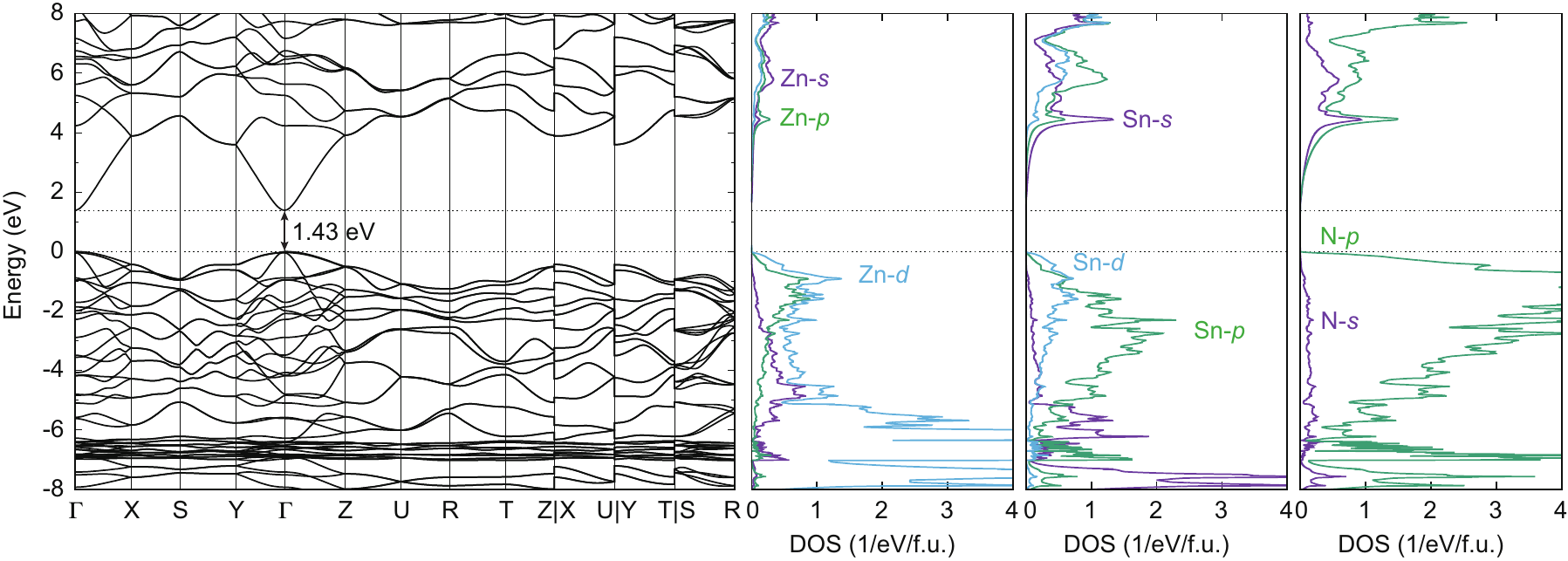}
  \caption{Band structure diagram and orbital-projected density of states (DOS) at each atomic site for the ordered orthorhombic ZnSnN$_2$ obtained using HSE06 at its theoretical lattice constants. The band path is based on Ref.~\cite{Hinuma2017140}. The energy zero is set at the VBM. Note that the DOS near the CBM is significantly small owing to the large band dispersion.}
  \label{band_dos_fig}
\end{figure}
\clearpage

\begin{figure}
  \centering
  \includegraphics[width=0.85\linewidth]{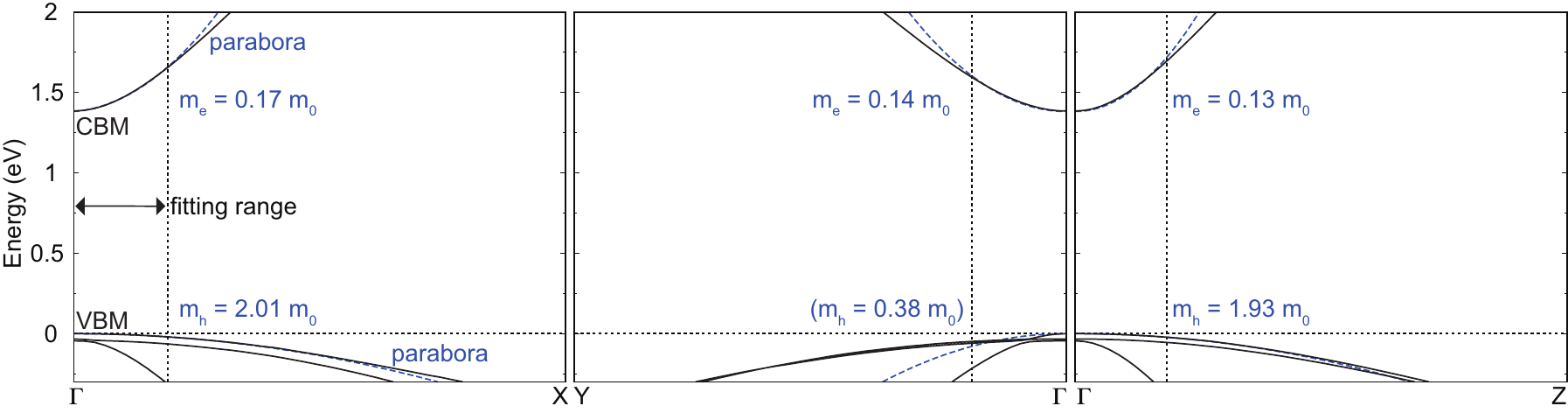}
  \caption{Electron ({\it m$_e$}) and heavy-hole ({\it m$_h$}) effective masses calculated through parabola fitting along the $\Gamma$-X, $\Gamma$-Y, and $\Gamma$-Z directions. The fitting range and the derived curves are also shown by dashed lines. The hole effective mass calculated from the split-off-like band is given in the parentheses as it does not significantly affect the averaged effective mass. These electron and heavy-hole effective masses are comparable or even superior to those of GaN ({\it m$_e$} =0.18{\it m$_0$} and {\it m$_h$} = 1.97{\it m$_0$}~\cite{Hinuma:2016ie}).}
  \label{effective_mass}
\end{figure}
\clearpage

\begin{figure}
  \centering
  \includegraphics[width=0.85\linewidth]{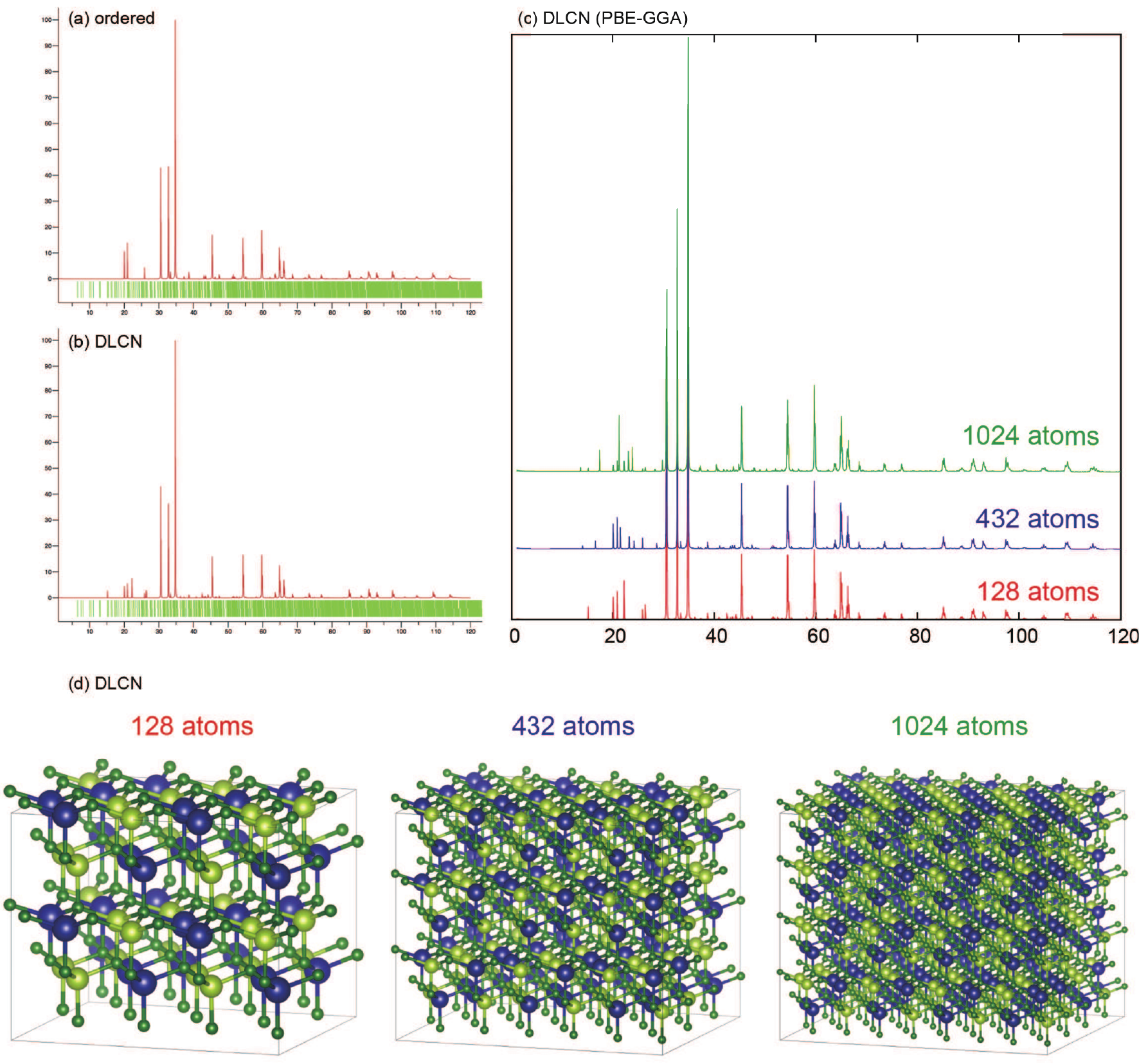}
  \caption{Simulated powder diffraction patterns for (a) the ordered structure and (b) 128-atom disordered structure with local charge neutrality (DLCN) based on the HSE06 structures. The VESTA code~\cite{JApplCryst.41.653} was used in the simulations. (c) Those for the DLCN models in different cell sizes optimized using PBEsol-GGA. The supercells used for (c) are shown in (d). Note that low angle peaks vary relatively largely depending on the configurations.}
  \label{xrd_fig}
\end{figure}
\clearpage

\begin{figure}
  \centering
  \includegraphics[width=0.85\linewidth]{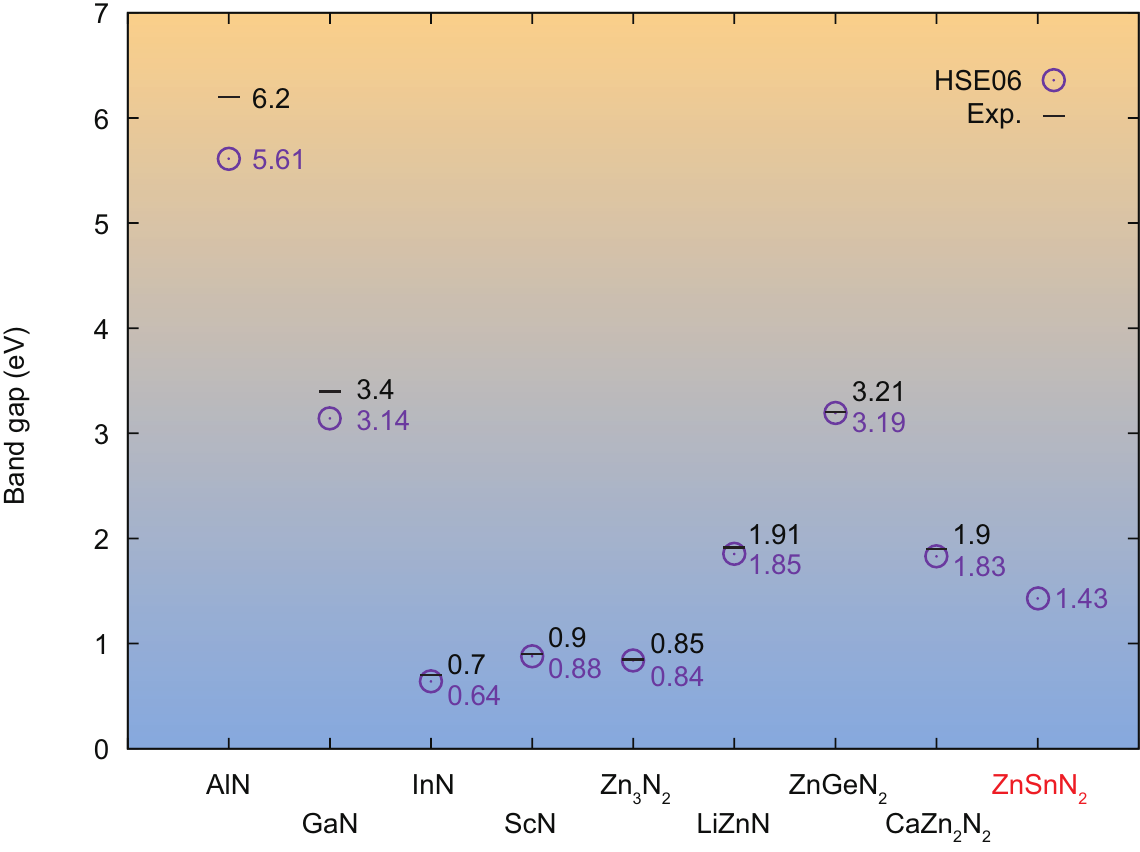}
  \caption{Experimental and HSE06 calculated band gaps of metal nitrides. Calculated values are denoted with circles and experimental values are shown with horizontal bars (see Refs.~\cite{doi:10.1063/1.1633965,1969_gan,doi:10.1063/1.1704853,PhysRevB.91.045104,Futsuhara:1998jw,Hinuma:2016ie}). The HSE06 functional accurately reproduces the band gaps up to $\sim$ 2 eV. Note that the BM shifts have been observed in InN, Zn$_3$N$_2$, and ScN, which lead to larger optical band gaps than the fundamental direct gaps. Thus, we show the smallest experimental values among those reported.}
  \label{bandgaps}
\end{figure}
\clearpage

\begin{figure}
  \centering
  \includegraphics[width=0.85\linewidth]{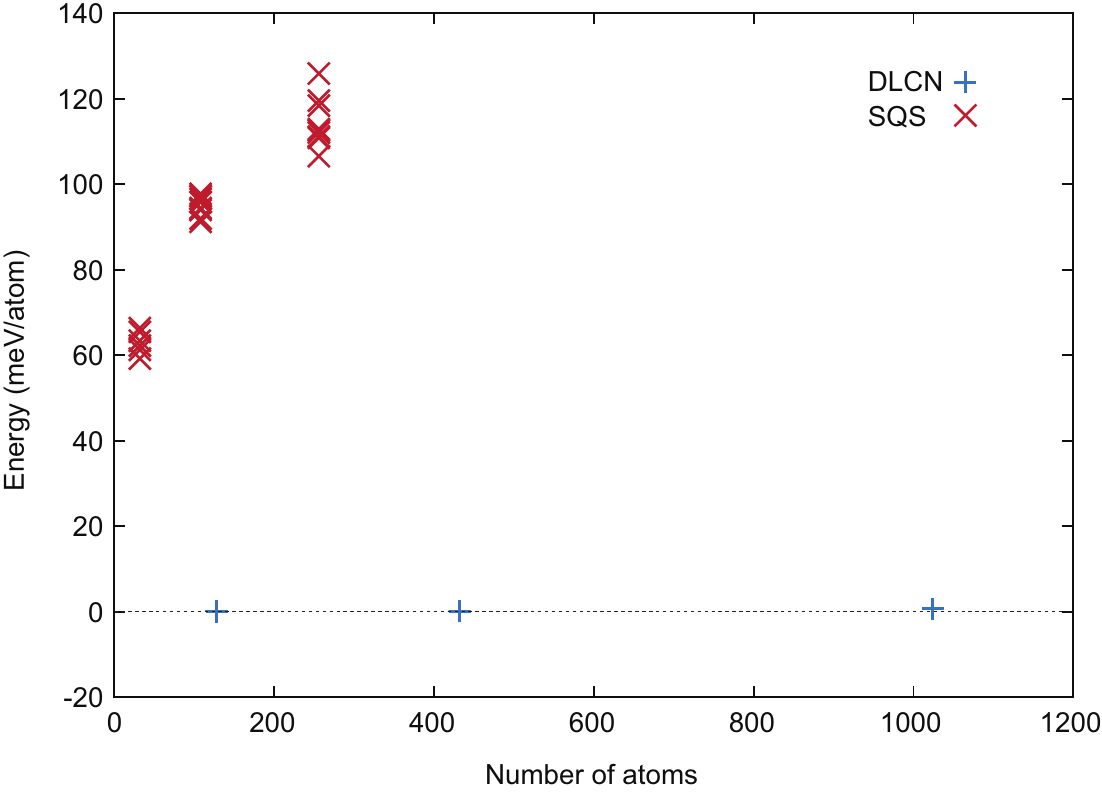}
  \caption{Relative total energies of special quasi-random structure (SQS) and DLCN models of ZnSnN$_2$ with different supercell sizes calculated using PBEsol-GGA. The energy zero is set to that of the ordered structure. The horizontal axis means the number of atoms in the supercells. Both DLCN and SQS models were generated by Monte Carlo simulated annealing (MCSA) using in-house and CLUPAN codes~\cite{PhysRevB.80.165122}, respectively. The SQS models were obtained by optimizing the correlation functions of pairs up to the 50th nearest neighbors to mimic the fully disordered state. To check the validity of the MCSA, we repeated the search of the SQS models ten times for each supercell. Lattice relaxation was allowed for all the structures. The energy differences indicate that the DLCN models have large advantages in energy than the SQS models with fully random cation disordering; the energies of the DLCN models differ by less than 1 meV/atom from that of the ordered model.}
  \label{opd_sqs_fig}
\end{figure}
\clearpage

\begin{figure}
  \centering
  \includegraphics[width=0.85\linewidth]{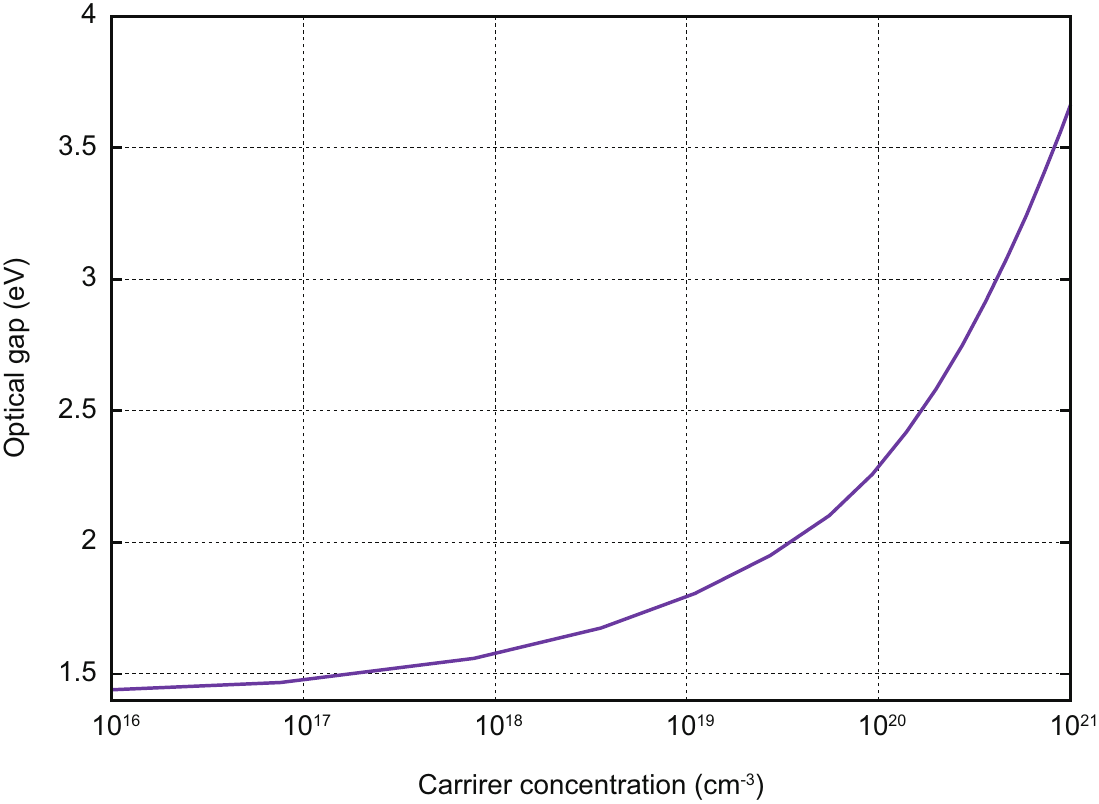}
  \caption{Optical gap as a function of the carrier-electron concentration for the ordered ZnSnN$_2$ calculated from the band structure and DOS using HSE06. The band dispersions of the valence and conduction bands were explicitly considered to evaluate the minimum vertical transition energy at each value of the carrier-electron concentration (see Ref.~\cite{PhysRevApplied.8.014015}).}
  \label{optical_gap_fig}
\end{figure}
\clearpage

\begin{figure}
  \centering
  \includegraphics[width=0.85\linewidth]{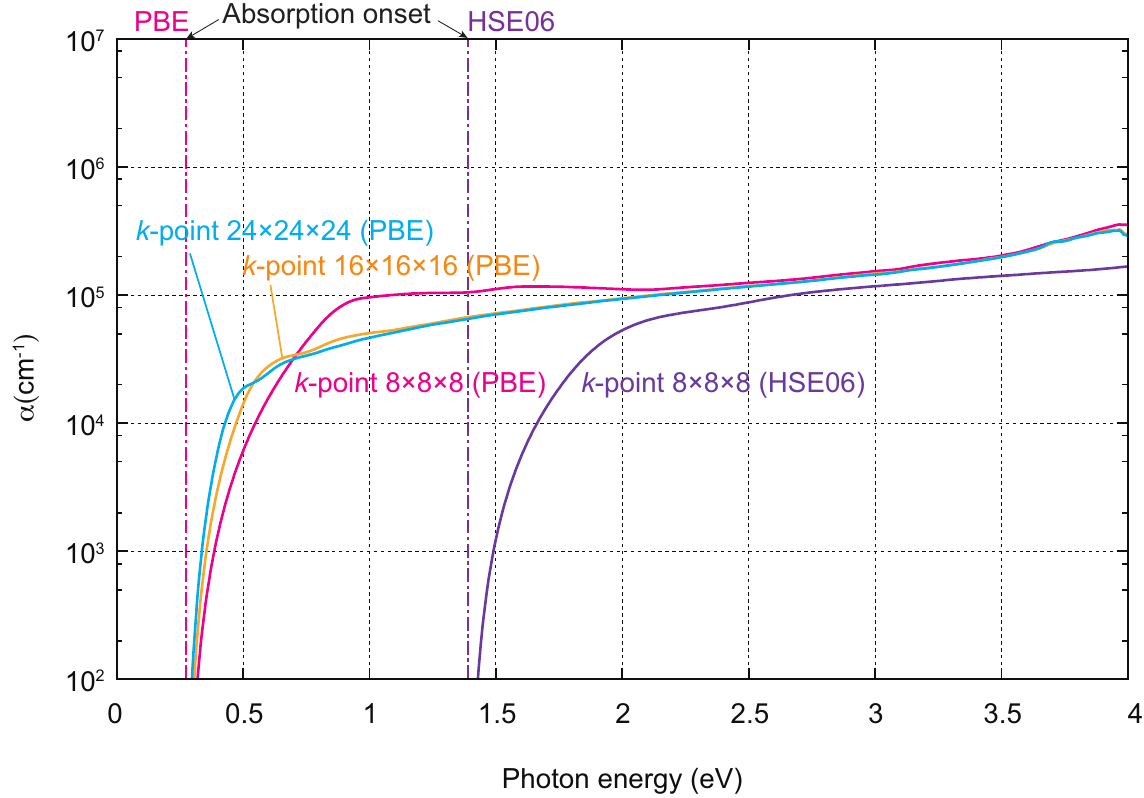}
  \caption{Absorption spectra for ordered ZnSnN$_2$ calculated using HSE06 and PBE-GGA~\cite{PhysRevLett.77.3865} with the 16-atom unit cell. The calculated band gaps with 8$\times$8$\times$8 $k$-point sampling are also shown by dashed-dotted lines. The HSE06 calculation was performed with 8$\times$8$\times$8 $k$-point sampling, whereas the PBE-GGA ones with 8$\times$8$\times$8, 16$\times$16$\times$16, and 24$\times$24$\times$24 $k$-point sampling; the lesser $k$-point set for the former is due to much higher computational cost of the HSE06 hybrid functional. The HSE06 result with 8$\times$8$\times$8 $k$-point sampling shows a sharp initial rise of the absorption spectrum near the band gap of 1.4 eV. Note, however, as inferred from the PBE results showing non-negligible $k$-point sampling dependencies, the HSE06 absorption coefficient would be increased near the onset when using well converged $k$-point sampling.}
  \label{absorption_fig}
\end{figure}
\clearpage

\begin{figure}
  \centering
  \includegraphics[width=0.85\linewidth]{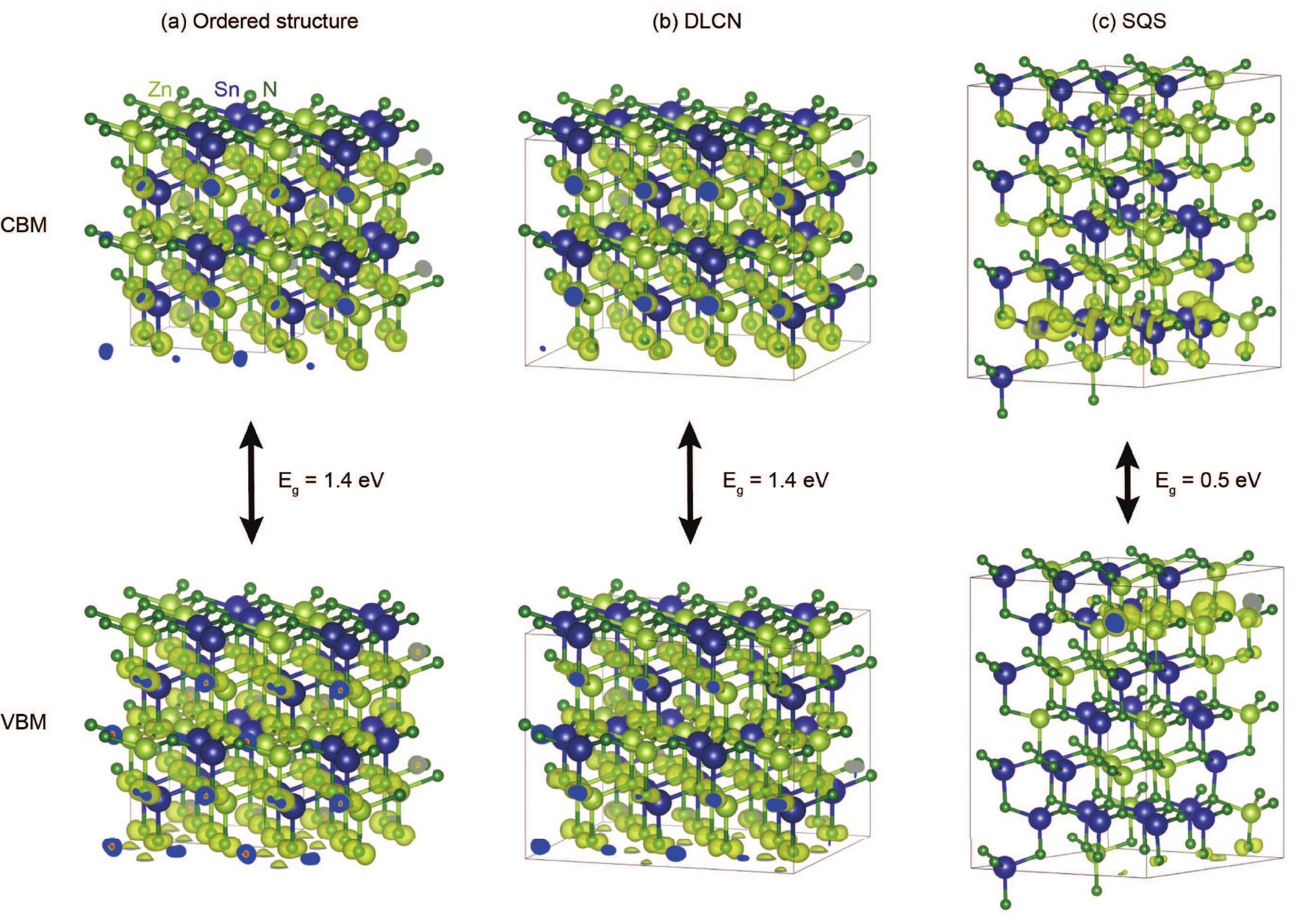}
  \caption{Isosurfaces of the squared wave functions at the VBM and CBM in (a) ordered, (b) DLCN, and (c) SQS ZnSnN$_2$. The band gap values are also shown. There is no significant difference in the distribution of the wave functions between the ordered and DLCN models. In contrast, the wave function at the VBM in the SQS model is largely localized at the N site with higher Zn coordination numbers, similarly to the Zn$_{\rm Sn}$ antisite defect (see the main text), while that at the CBM is localized at the Sn site. The VESTA code~\cite{JApplCryst.41.653} was used for visualization.}
  \label{charge_fig}
\end{figure}
\clearpage

\begin{figure}
  \centering
  \includegraphics[width=0.85\linewidth]{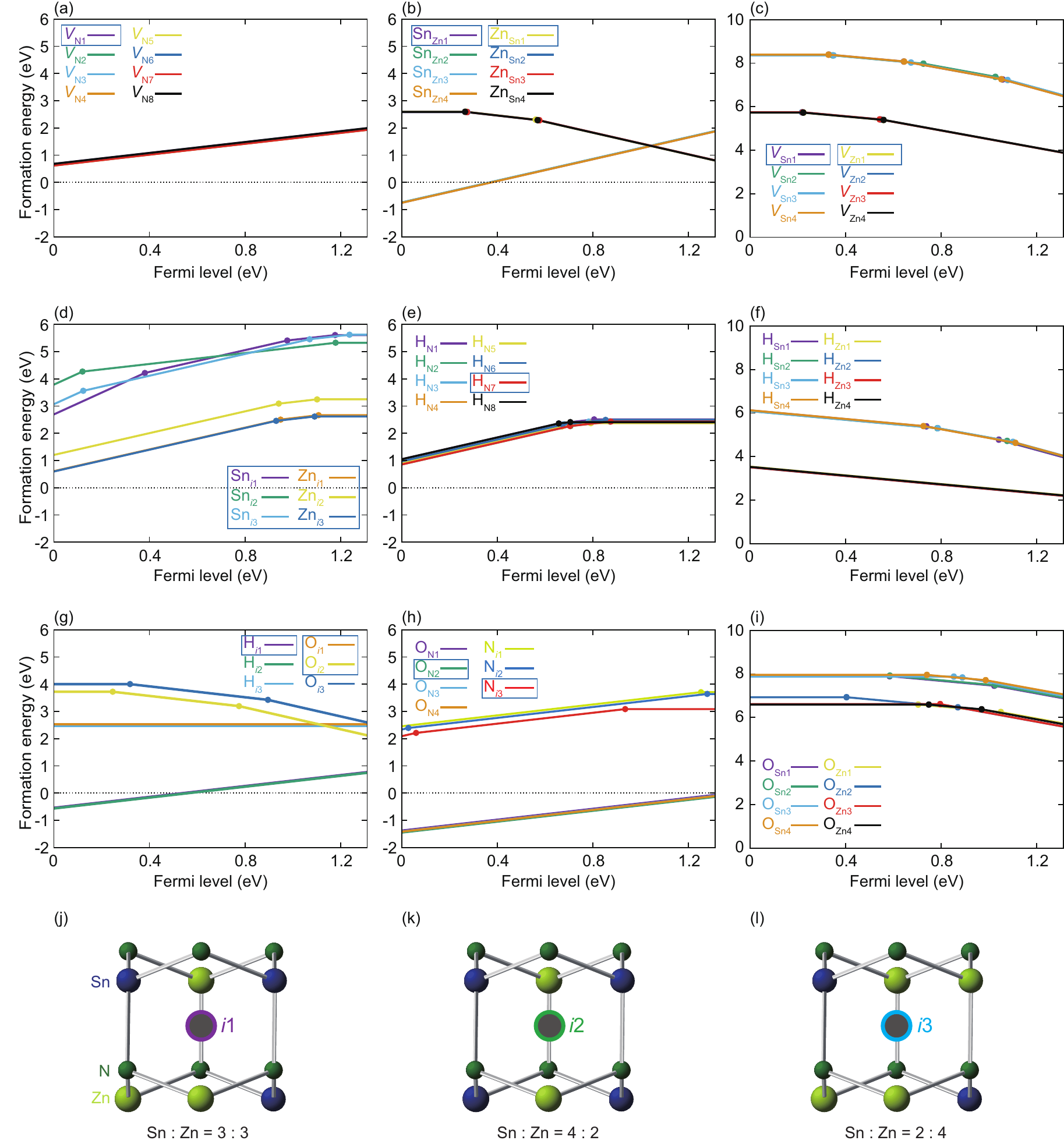}
  \caption{(a--i) Site dependencies of the defect-formation energies in the DLCN model shown in Figure 1(b) in the main text as a function of the Fermi level and (j, k, l) local structures near these interstitial sites considered. Here, PBEsol-GGA was used in conjunction with the 2$\times$2$\times$2 Monkhorst-Pack (MP) $k$-point mesh. Errors associated with the finite supercell size were corrected for all the defects using the extended FNV correction scheme. One can see that site dependences are negligibly small, especially for defects with lower formation energies. Relatively large variations are seen in interstitial-type defects, presumably because of the difference in neighboring cation configurations as shown in (j, k, l); the difference in the neighboring Sn:Zn ratios results in different electrostatic effects on the interstitial-type defects. H$_X$ and O$_X$ sites ($X$= Zn, Sn) are not considered in HSE06 calculations because their energies are much higher than those of H$_i$ and O$_{\rm N}$. Configurations enclosed in squares were adopted for the HSE06 calculations in the main text. Spin polarization was not considered here in order to reduce the computational costs.}
  \label{site_fig}
\end{figure}
\clearpage

\begin{figure}
  \includegraphics[width=1\linewidth]{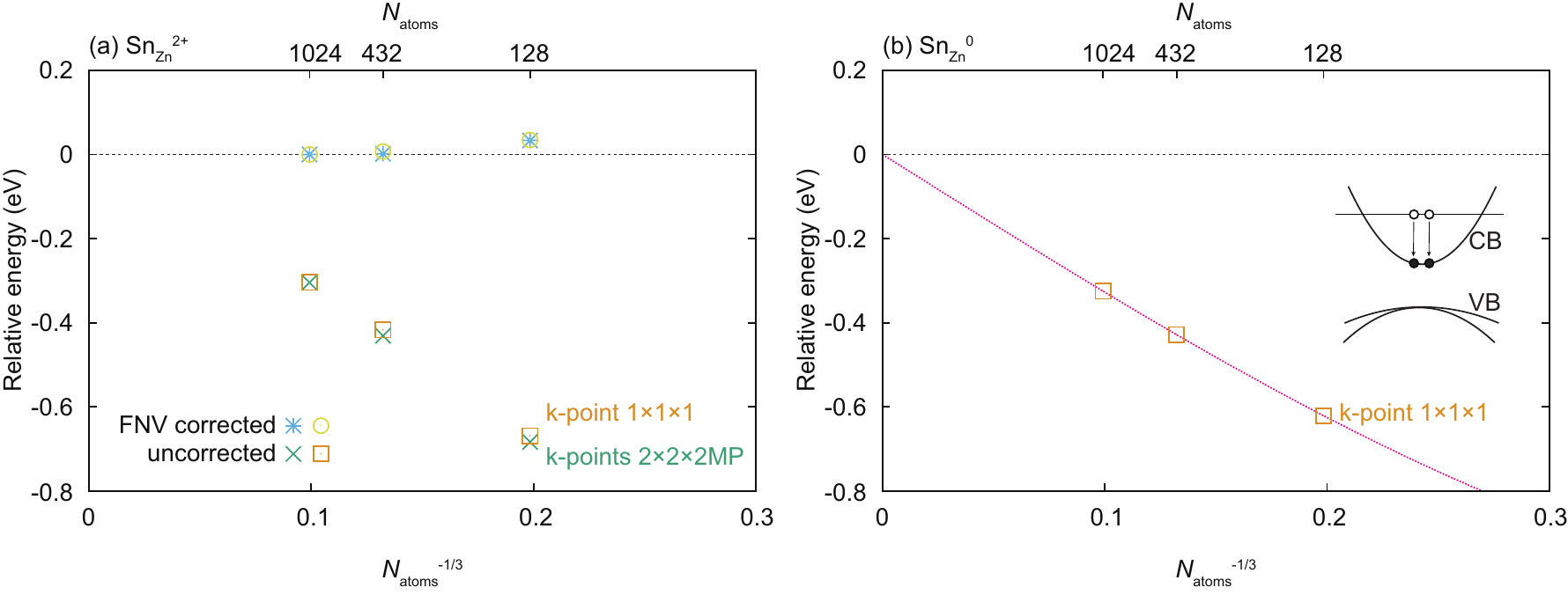}
  \caption{Formation energies of (a) Sn$_{\rm Zn}^{2+}$ and (b) Sn$_{\rm Zn}^{0}$ in the ordered ZnSnN$_2$ as a function of the supercell size using PBEsol-GGA. $N$$_{\rm atom}$ is the number of atoms in the supercell. In the case of Sn$_{\rm Zn}^{2+}$, corrected formation energies using the extended FNV method are also shown. The energy zero is set at the corrected energy calculated using the largest 1024-atom supercell. The estimated error in the 128-atom supercell is 0.72 eV, but it is reduced to less than 0.05 eV when the FNV corrections are applied. In the case of Sn$_{\rm Zn}^{0}$, the supercells contain a doubly occupied hydrogenic donor state whose electrons are released from a defect state as shown by the schematic illustration. One can see a large cell size dependence, although the charge state is neutral. This is mostly attributed to the interaction between the doubly ionized Sn$_{\rm Zn}$, its periodic images, and hydrogenic donor electrons that spread throughout the supercell and behave like the background charge. See Refs.~\cite{PhysRevB.77.245202,PhysRevApplied.8.014015} for more details.}
  \label{cell_kpt_fig}
\end{figure}
\clearpage

\begin{figure}
  \centering
  \includegraphics[width=0.85\linewidth]{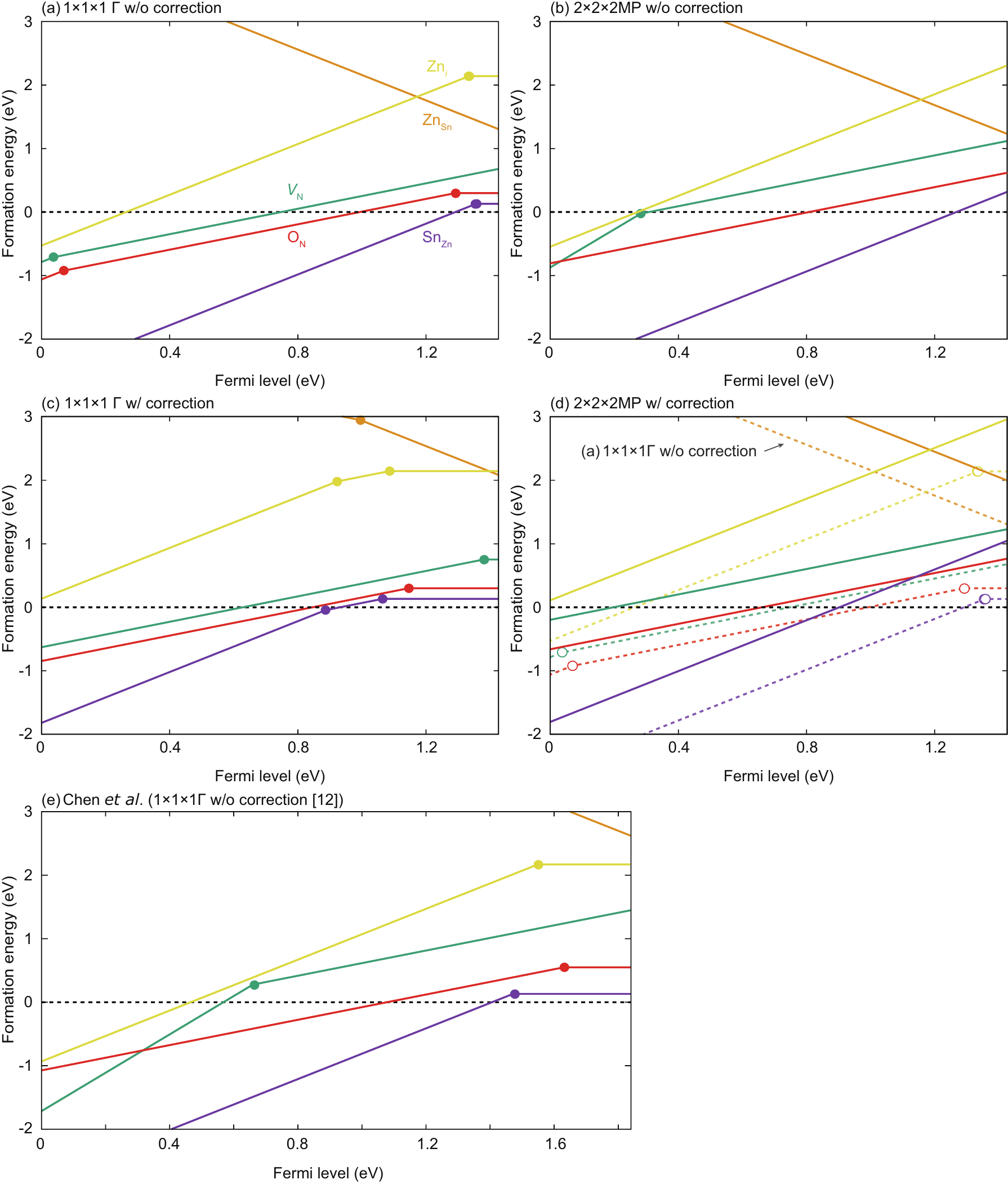}
  \caption{Defect-formation energies as a function of the Fermi level in the ordered ZnSnN$_2$ (a, b) without (w/o) and (c, d) with (w/) finite cell size corrections. (a, c) The $\Gamma$-point and (b, d) the 2$\times$2$\times$2 MP $k$-point mesh were used for reciprocal space sampling. The cell size corrections were performed using the extended FNV correction scheme. One can see the importance of the cell-size corrections and $k$-point sampling. (e) The formation energies calculated by Chen et al. reproduced from Ref.~\cite{chen_2014}. They used the same calculation condition with (a), namely no finite-cell size corrections and $\Gamma$-point only $k$-point sampling, but the mixing parameter of HSE was increased from 0.25 to 0.31, resulting in a larger band gap.}
  \label{ttl_fig}
\end{figure}
\clearpage

\begin{figure}
  \centering
  \includegraphics[width=0.85\linewidth]{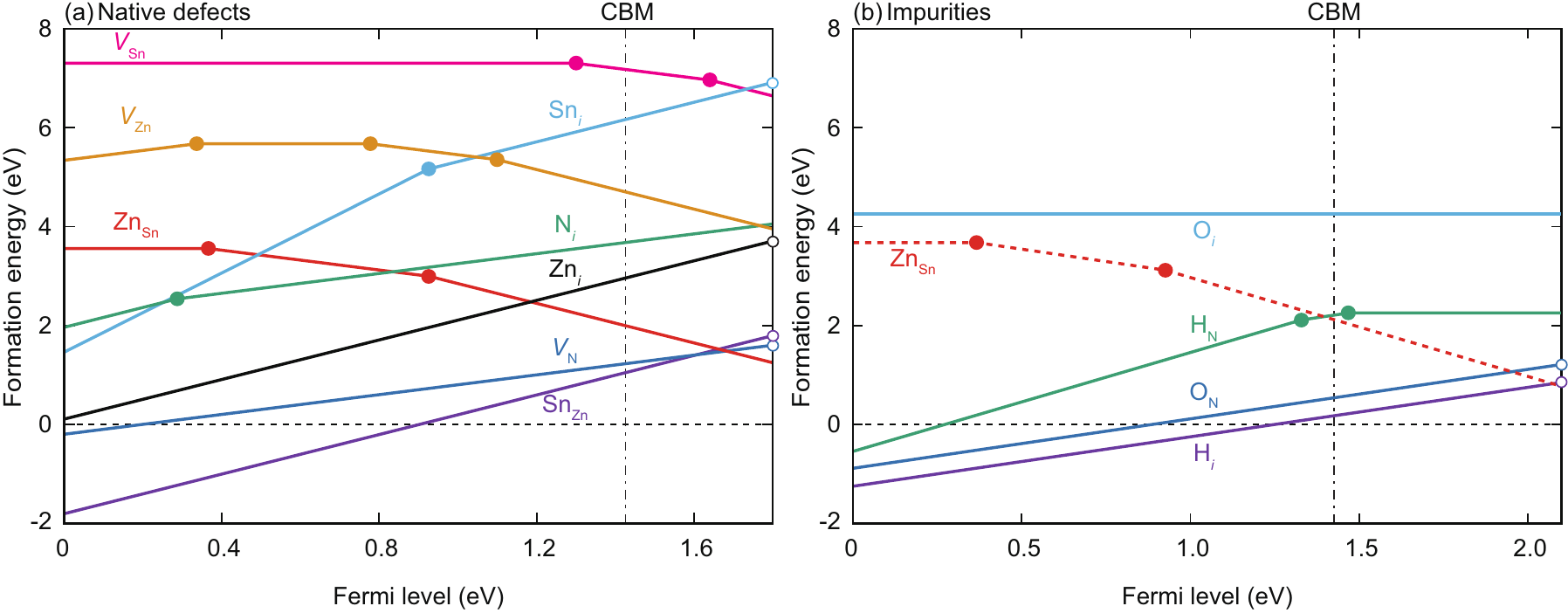}
  \caption{Formation energies of (a) native defects and (b) impurities in the ordered ZnSnN$_2$ as a function of the Fermi level. In (b), the formation energy of the dominant acceptor Zn$_{\rm Sn}$ among the native defects is also shown for comparison. The upper limit of the Fermi level is extended up to (a) 1.8 eV and (b) 2.1 eV with respect to the VBM, respectively, so that the intersections between the formation energies of the dominant donors and acceptors are visible. The transition levels above the CBM are calculated properly when the $k$-point mesh used for the defect calculations does not sample the CBM. In this study, we used the 2$\times$2$\times$2 MP $k$-point mesh, in which the lowest unoccupied state in the perfect supercell is located at 2.34 eV above the VBM (see Ref.~\cite{PhysRevApplied.8.014015} for details).}
  \label{extended_fig}
\end{figure}
\clearpage

\begin{figure}
  \centering
  \includegraphics[width=0.85\linewidth]{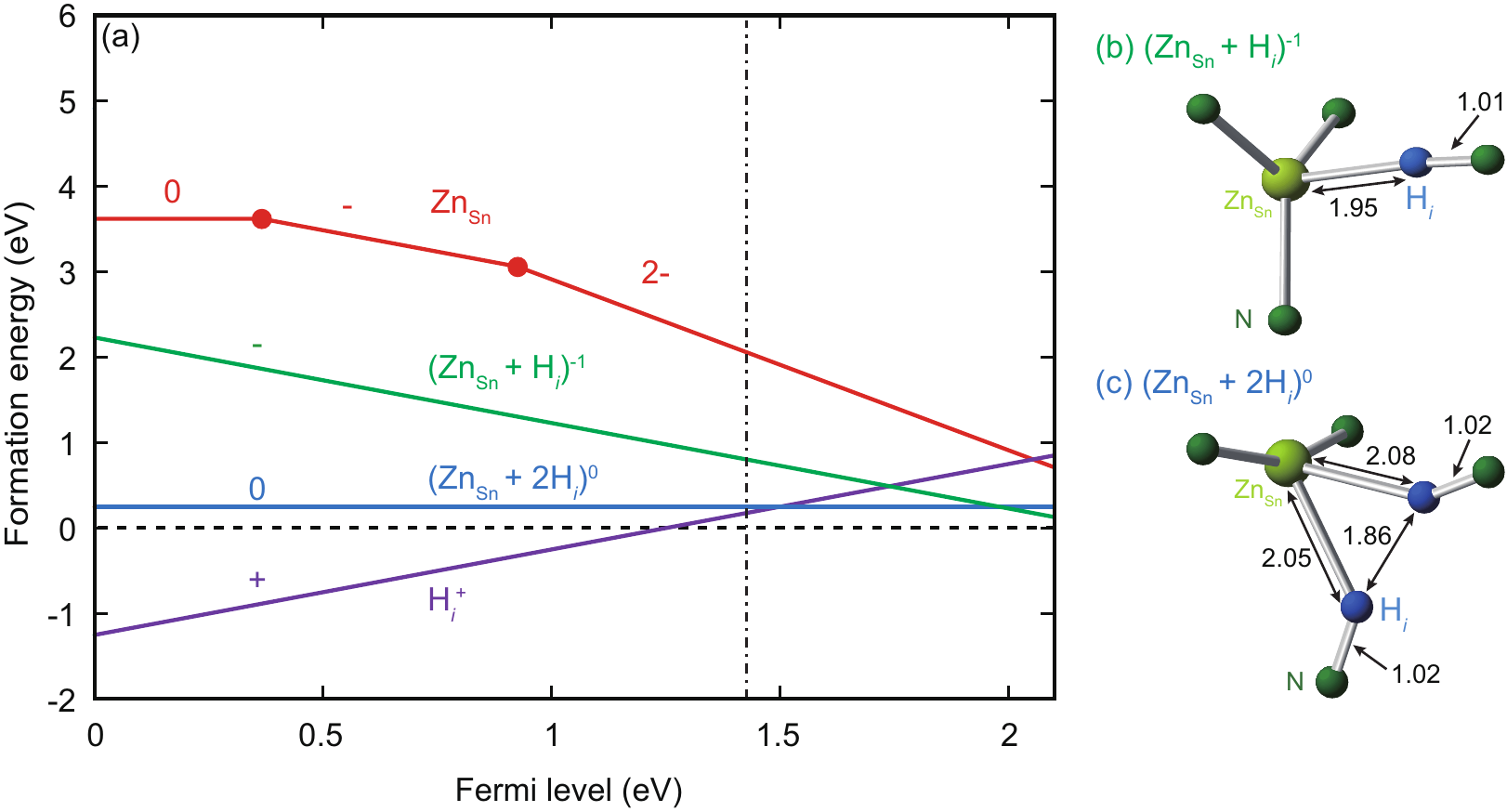}
  \caption{(a) Formation energies of H$_i$ and Zn$_{\rm Sn}$ and their complex defects (Zn$_{\rm Sn}$ + H$_i$)$^{-}$ and (Zn$_{\rm Sn}$ + 2H$_i$)$^{0}$ calculated using HSE06. Relaxed atomic structures of (Zn$_{\rm Sn}$ + H$_i$)$^{-}$ and (Zn$_{\rm Sn}$ + 2H$_i$)$^{0}$ are shown in (b) and (c), respectively. The distances between atoms indicated by arrows in (b) and (c) are shown in \AA. Note that only one configuration for each complex defect is considered. Complexing with hydrogen is exothermic and drastically decreases the formation energy of acceptor Zn$_{\rm Sn}$. See main text for details.}
  \label{extended_fig}
\end{figure}
\clearpage

\begin{figure}
  \centering
  \includegraphics[width=0.85\linewidth]{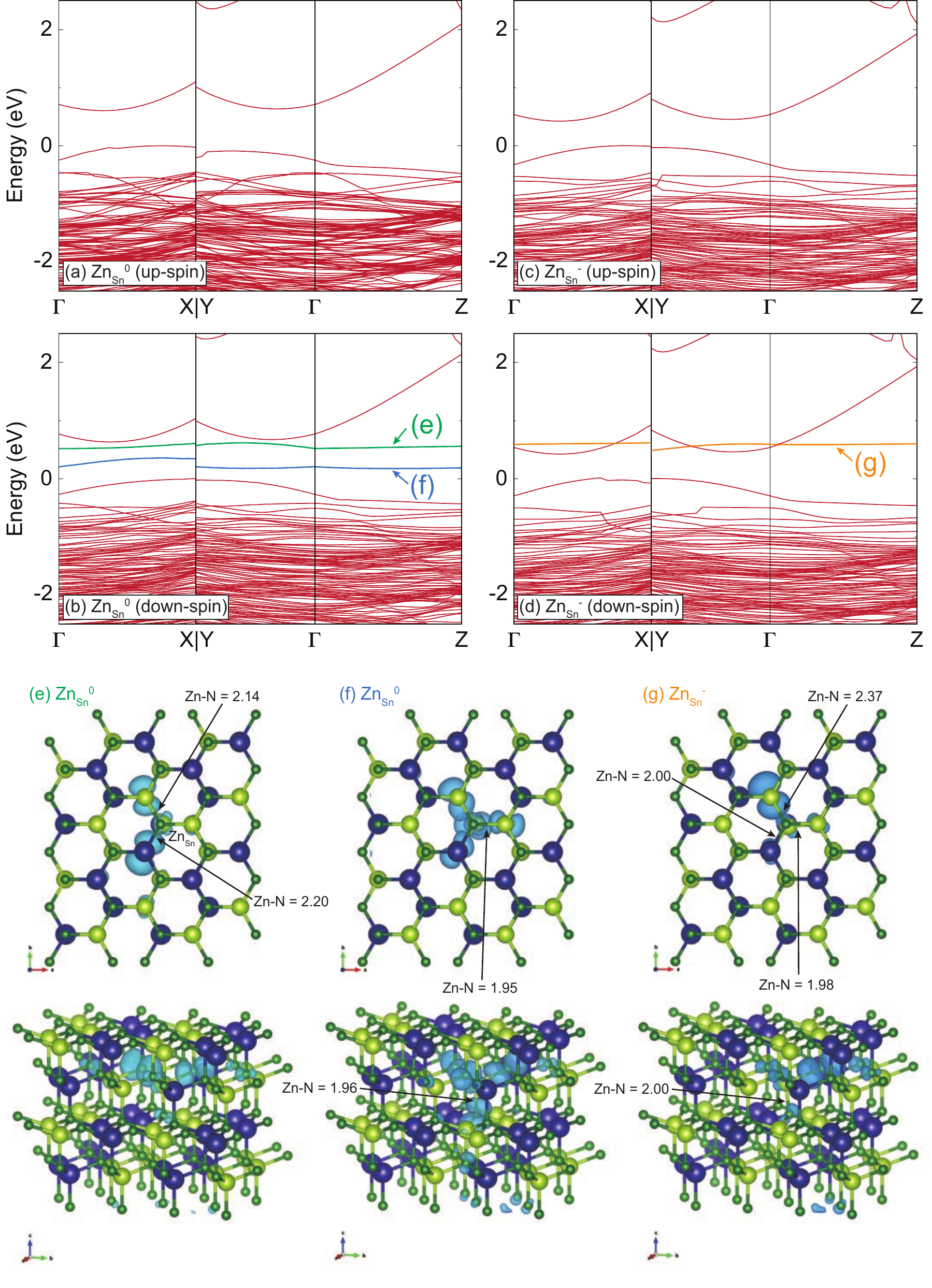}
	\caption{(a--d) Band structures for the up-spin and down-spin channels of Zn$_{\rm Sn}^{0}$ and Zn$_{\rm Sn}^{-}$ in the 128-atom supercell of the ordered model. (e--g) Squared wave functions of the bands highlighted in Figures (b) and (d), indicating localized defect states. The distances of Zn$_{\rm Sn}$-N are indicated by arrows in \AA. Note that the Zn$_{\rm Sn}^{2-}$-N distances are 2.04--2.07 \AA. As discussed in the main text, two holes in the Zn$_{\rm Sn}^{0}$ model are located in the down-spin channel within the band gap, while one hole in the Zn$_{\rm Sn}^{-}$ model is in the same channel but slightly above the CBM. The deep levels exist as hole polaronic states captured by the neighboring N atoms with accompanying outward relaxations as seen in (e--g).}
  \label{extended_fig}
\end{figure}
\clearpage

\begin{figure}
  \centering
  \includegraphics[width=0.85\linewidth]{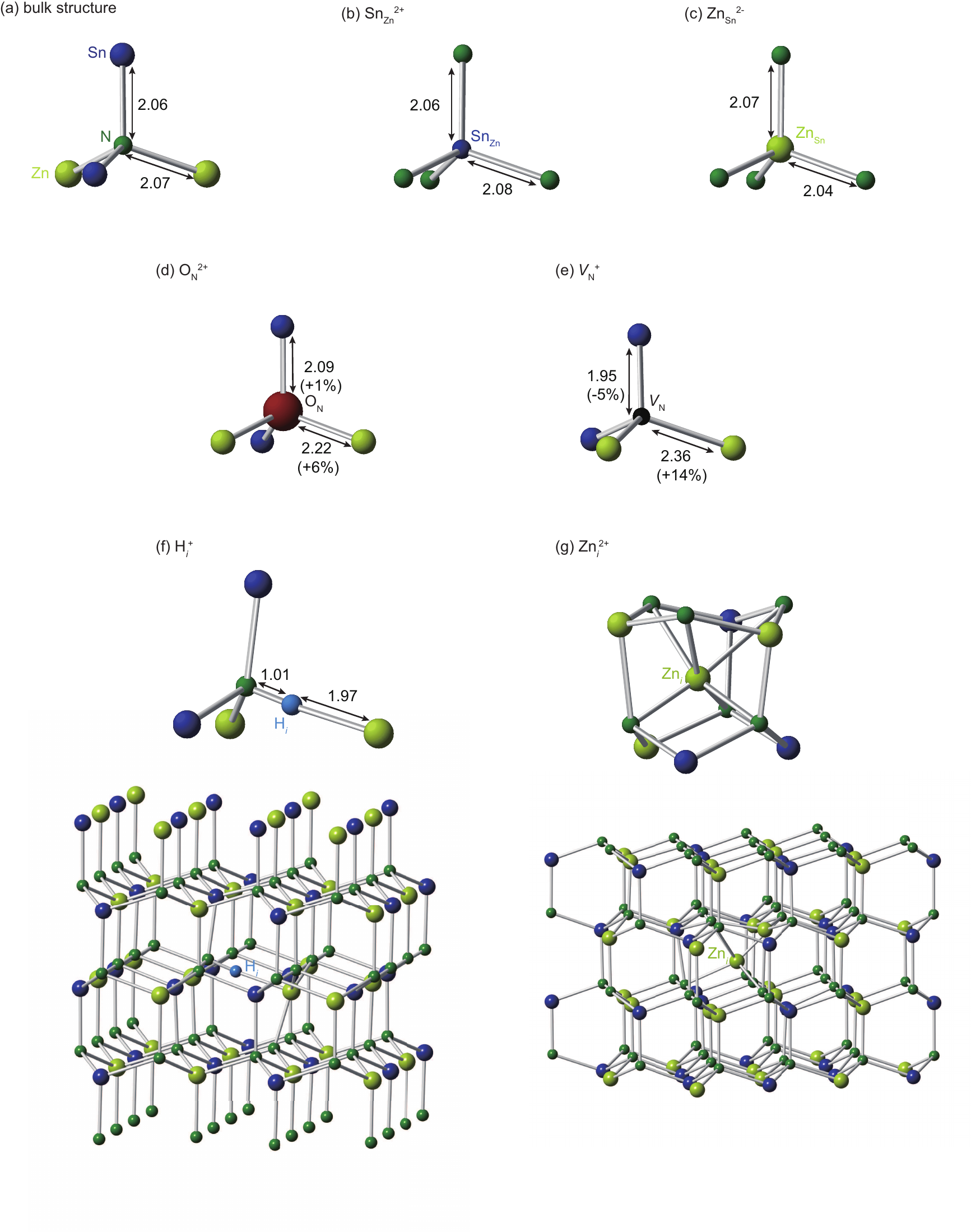}
  \caption{Local defect structures in the ordered ZnSnN$_2$. The distances between atoms are shown in \AA. For comparison, the tetrahedral configuration around an N atom in the ordered ZnSnN$_2$ is shown in (a). Sn$_{\rm Zn}^{2+}$, Zn$_{\rm Sn}^{2-}$, and O$_{\rm N}^{+}$ in (b), (c), and (d), respectively, show relatively small displacements of coordinating atoms probably because of similar ionic radii between Zn and Sn and between N and O. $V_{\rm N}^{+}$ in (e) shows inward relaxations of Sn atoms with an occupied Sn-Sn bond and outward relaxations of Zn atoms. H$_i^{+}$ in (f) is located in between N and Zn with forming a strong N-H bond. Zn$_i^{2+}$ in (g) is slightly off-centered from the octahedral site. Larger areas are also depicted for (f) H$_i^{+}$ and (g) Zn$_i^{2+}$.}
  \label{defect_fig}
\end{figure}
\clearpage

\begin{figure}
  \centering
  \includegraphics[width=0.65\linewidth]{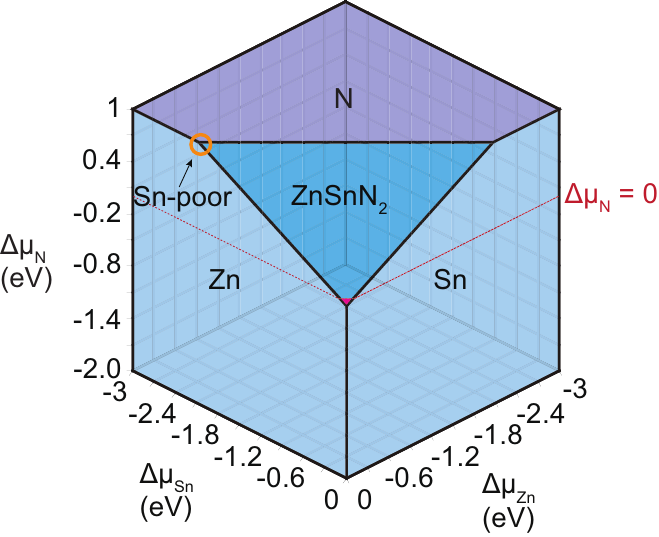}
  \caption{Chemical potential diagram that shows the condition considered to achieve the low carrier-electron concentration in ZnSnN$_2$. As discussed in the main text, the nitrogen chemical potential $\Delta \mu_{\rm N}$ is increased by 1 eV from the energy of the N$_2$ molecule per atom. The equilibrium point used for the calculations in Figure 3 in the main text is designated by an open circle, corresponding to the Sn-poor (Zn-rich) condition.}
  \label{hull_fig}
\end{figure}
\clearpage

\begin{figure}
  \centering
  \includegraphics[width=0.85\linewidth]{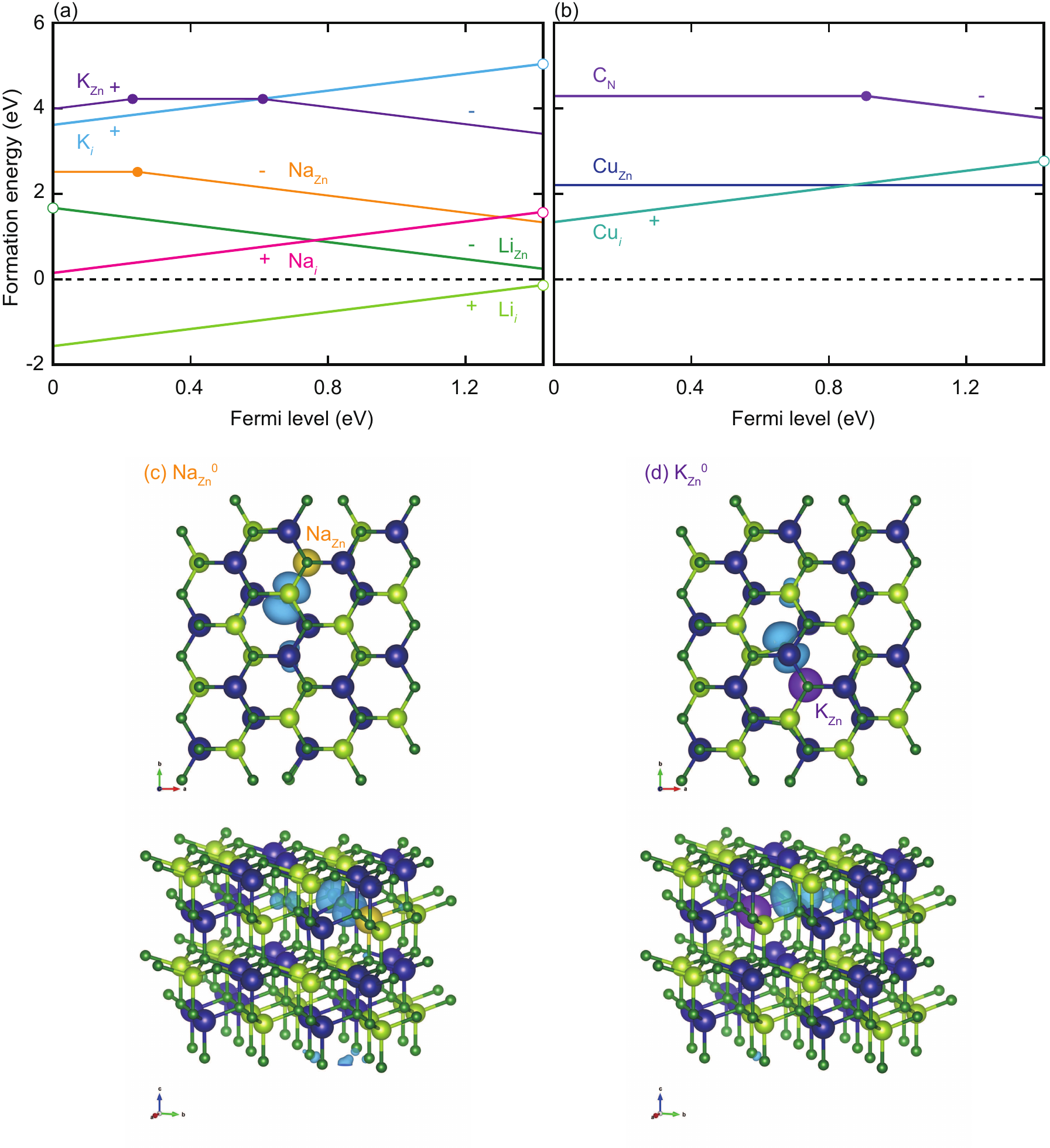}
	\caption{Formation energies of substitutional and interstitial (a) Li, Na, and K, and (b) Cu and C considered as acceptor dopants in the ordered ZnSnN$_2$ at the Sn-poor (Zn-rich) condition in Figure S16. (c, d) Squared wave functions of the hole trapping states in Na$_{\rm Zn}^{0}$, and K$_{\rm Zn}^{0}$ models. All the monovalent cations (Li, Na, K, and Cu) favor the interstitial site in the $p$-type regime, and Cu$_{\rm Zn}$ and C$_{\rm N}$ show extremely deep acceptor levels, indicating difficulty of $p$-type doping with these dopants. It is also noteworthy in view of electronic structure that Na$_{\rm Zn}$ and K$_{\rm Zn}$ show similar polaronic behaviors to Zn$_{\rm Sn}$; they exist as hole polarons locating at the N sites near the defects, as seen in (c) and (d). In the cases of C$_{\rm N}$ and Cu$_{\rm Zn}$, on the other hand, a hole is localized at the C-$p$ state and Cu-3$d$ state, respectively.}
  \label{acceptor_fig}
\end{figure}
\clearpage
